\documentclass[twocolumn,prl,preprintnumbers,amsmath,amssymb]{revtex4}

\usepackage{graphicx}
\usepackage{times}
\usepackage{amsmath}
\usepackage{amsfonts}
\usepackage{amssymb}
\usepackage{inputenc}

\newcommand{\m}[1]{\ensuremath{\mathrm{#1}}}

\usepackage{color}
\definecolor{red}{rgb}{1,0,0}
\definecolor{blue}{rgb}{0,0,1}
\definecolor{black}{rgb}{0,0,0}

\begin{document}

\title{Identifying optical signatures of momentum-dark excitons in transition metal dichalcogenide monolayers}

\author{Jessica Lindlau$^{1}$, Cedric Robert$^{2}$, Victor Funk$^{1}$, Jonathan F\"orste$^{1}$, Michael F\"org$^{1}$, L\'{e}o Colombier$^{1}$, Andre Neumann$^{1}$, Emmanuel Courtade$^{2}$, Shivangi Shree$^{2}$, Marco Manca$^{2}$, Takashi Taniguchi$^{3}$, Kenji Watanabe$^{3}$, Mikhail M. Glazov$^{4}$, Xavier Marie$^{2}$, Bernhard Urbaszek$^{2}$, and Alexander
H\"ogele$^{1}$}

\affiliation{$^1$Fakult\"at f\"ur Physik, Munich Quantum Center,
and Center for NanoScience (CeNS),
Ludwig-Maximilians-Universit\"at M\"unchen,
Geschwister-Scholl-Platz 1, 80539 M\"unchen, Germany}

\affiliation{$^2$Universit\'{e} de Toulouse, INSA-CNRS-UPS, LPCNO,
135 Ave. de Rangueil, 31077 Toulouse, France}

\affiliation{$^3$National Institute for Materials Science,
Tsukuba, Ibaraki 305-0044, Japan}

\affiliation{$^4$Ioffe Institute, 26 Polytechnicheskaya, 194021
St. Petersburg, Russia}

\date{\today}

\begin{abstract}
Transition metal dichalcogenide (TMD) monolayers (MLs) exhibit
rich photoluminescence spectra associated with interband optical
transitions of direct-gap semiconductors. Upon absorption of
photons, direct excitons with zero center-of-mass momentum are
formed by photo-excited electrons in the conduction band and the
respective unoccupied states in the valence band of the same
valley. Different spin configurations of such momentum-direct
excitons as well as their charged counterparts provide a powerful
platform for spin-valley and microcavity physics in
two-dimensional materials. The corresponding spectral signatures,
however, are insufficient to explain the main characteristic peaks
observed in the photoluminescence spectra of ML TMDs on the basis
of momentum-\textit{direct} excitons alone. Here, we show that the
notion of momentum-\textit{indirect} excitons is important
for the understanding of the versatile photoluminescence
features. Taking into account phonon-assisted radiative
recombination pathways for electrons and holes from dissimilar
valleys, we interpret unidentified peaks in the emission spectra
as acoustic and optical phonon sidebands of momentum-dark
excitons. Our approach
will facilitate the interpretation of optical,
valley and spin phenomena in TMDs arising from bright and dark
exciton manifolds.
\end{abstract}

\maketitle

Monolayer (ML) transition metal dichalcogenides (TMDs) are
material representatives of a broader class of atomically thin
direct-gap semiconductors \cite{Splendiani2010,Mak2010} pivotal
for the realization of van der Waals heterostructures and devices
with novel functionality \cite{Geim2013,Novoselov2016,Mak2016a}.
They feature strong optical transitions promoted by excitons
\cite{Chernikov2014,He2014,Ye2014,Wang2017b} and, paired with
valley-selective excitation \cite{Xiao2012}, manipulation
\cite{Kim2014,Sie2015,Wang2016,Ye2017,Sie2017} and detection
\cite{Cao2012,Mak2012,Zeng2012,Jones2013} schemes, represent
viable resources for opto-valleytronic applications
\cite{Yao2008,Xu2014,Neumann2017}. While optical transitions of lowest-energy
excitons in molybdenum-based MoSe$_2$, MoS$_2$ or MoTe$_2$ MLs is
spin-allowed, the exciton ground state of ML tungsten
dichalcogenides WSe$_2$ and WS$_2$ is spin-forbidden
\cite{Zhang2015,Wang2015,Withers2015,Molas2017,Zhang2017,Zhou2017,WangMak2017,Wang2017a}.
This striking difference stems from a reversed energetic ordering
of spin-polarized conduction sub-bands in molybdenum and tungsten
dichalcogenide MLs
\cite{Liu2013,Kosmider2013a,Kosmider2013b,Kormanyos2014,Kormanyos2015}.

While early photoluminescence (PL) spectroscopy studies have
established elementary signatures of bright excitons in neutral
\cite{Chernikov2014,He2014,Ye2014} and charged TMD MLs
\cite{Mak2013,Ross2013,Sidler2017}, the emission from spin-forbidden
excitons has been identified only recently
\cite{Zhang2015,Molas2017,Zhang2017,Zhou2017,WangMak2017,Wang2017a}.
The observations of lowest-lying momentum-bright yet spin-forbidden
states in tungsten dichalcogenide MLs explain some of the
differences between the rich structure in the PL spectra of
tungsten-based MLs and the rather simple one- or two-peak PL of ML
molybdenum dichalcogenides \cite{Cadiz2017}. Some of the main PL
peaks that can be more intense than the bright exciton, however,
have escaped unambiguous assignment and are thus commonly
attributed to defect-localized excitons. Moreover, unequivocal
deconvolution of individual PL contributions from neutral and
charged excitons has been compromised by the lack of control over
the charge doping level in most samples and impeded further by the
conspiracy of similar energy scales of optical phonons
\cite{Jin2014,Dery2015a} and trion binding energies
\cite{Ross2013,Jones2013,Jones2015}. Here, we propose a unifying
explanation for unidentified PL features in the spectra of TMD MLs
by expanding the realm of momentum-direct excitons with their
momentum-indirect counterparts. This analysis benefits from the
greatly improved optical quality of TMD MLs encapsulated in hBN
\cite{Dean:2010a,WangMak2017,Ajayi2017,Cadiz2017,Jin:2016a} in
charge tunable structures.

\begin{figure*}[t]
\begin{center}
\includegraphics[scale=1.07]{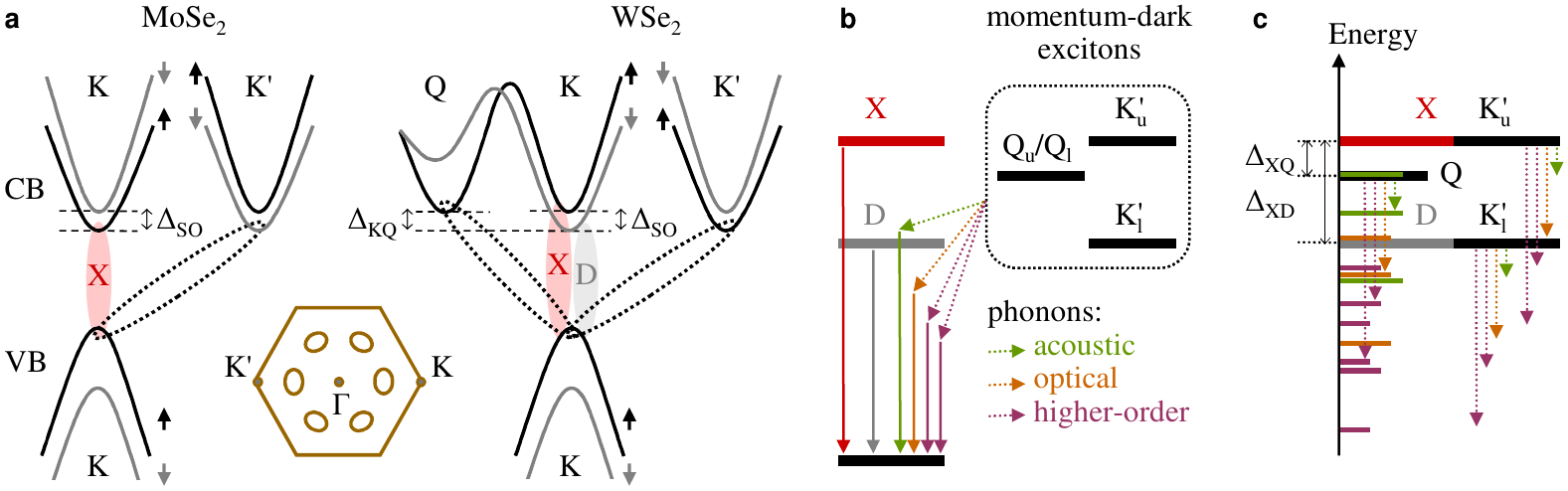}
\caption{\textbf{Basic concepts for the spectral decomposition of
photoluminescence from monolayer transition metal
dichalcogenides.} \textbf{a},~Schematic band structure in the
first Brillouin zone (inset with high-symmetry points
$\m{\Gamma}$, $\m{K}$ and $\m{K'}$, and six inequivalent
$\m{Q}$-pockets) of molybdenum (left) and tungsten (right)
dichalcogenide monolayers. Conduction band (CB) and valence band
(VB) with spin-up and spin-down electron sub-bands (shown in black
and grey, respectively), spin-orbit splitting $\Delta_{\m{SO}}$,
and the energy separation $\Delta_{\m{KQ}}$ between the conduction
band minima at $\m{K}$ and $\m{Q}$. Momentum-direct spin-bright
and spin-forbidden excitons ($\m{X}$ and $\m{D}$, indicated
by ellipses shaded in red and grey) are formed by electrons and
unoccupied states in the $\m{K}$-valley. Momentum-dark excitons
(dashed ellipses) with the empty state at $\m{K}$ can be formed
with electrons at $\m{Q}$ or $\m{K'}$. \textbf{b} and
\textbf{c}, With this realm of direct and phonon-assisted
radiative processes, we construct a conceptual PL spectrum as the
sum of the zero-phonon lines (ZPLs) of $\m{X}$ and $\m{D}$
excitons (red and grey bars) and the phonon side-bands of
momentum-dark excitons (colored bars). The energy position of the
latter (black bars) can be reconstructed from their respective
phonon replicas (green and orange bars) by considering in-plane
transverse acoustic (TA) and longitudinal acoustic (LA) phonons as
well as in-plane TO(E$'$), LO(E$'$) and out-of-plane $\m{A}_1$
optical phonon modes, and sidebands resulting from higher order
decay processes (purple bars) assisted by combinations of multiple
phonons.
}
\label{fig1}
\end{center}
\end{figure*}
\textbf{Signatures of direct and indirect excitons in the emission
spectra}. First we describe our approach to decompose the optical
spectra of ML TMDs involving momentum-dark exciton contributions.
Subsequently, we demonstrate how this simple approach can be
applied to reproduce the most prominent spectral features observed
in PL. The basic understanding of the optical phenomena in TMD MLs
derives from the single-particle band structure shown
schematically in Fig.~\ref{fig1}a. The conduction band (CB) and
valence band (VB) feature spin-polarized sub-bands with energy
splittings $\Delta_{\m{SO}}$ at $\m{K}$ and $\m{K'}$ points of the
first Brillouin zone. The VB spin-orbit splitting of a few hundred
of meV as estimated from first-principles calculations
\cite{Liu2013,Kosmider2013a,Kosmider2013b,Kormanyos2015} and
determined experimentally \cite{Zhang2014a,Wang2014} is contrasted
by a much smaller CB splitting $\Delta_{\m{SO}}$ on the order of a
few to a few tens of meV
\cite{Liu2013,Kosmider2013a,Kosmider2013b,Kormanyos2015}. In
addition to the $\m{K}$ and $\m{K'}$ valleys, the CB of TMD MLs
exhibits local minima at six non-equivalent $\m{Q}$-pockets
related pairwise by time-reversal symmetry
\cite{Zhao2013,Liu2015}. Depending on the specific material and
the details of calculations, the $\m{Q}$-valley band-edges can be
as far as $\Delta_{\m{KQ}} \simeq 160$~meV above the CB minimum as
in MoSe$_2$, or in the range of $\sim 0 - 80$~meV in
tungsten-based MLs \cite{Jin2014,Kormanyos2015,Zhang2015a}.\\

With this single-particle picture in mind we interpret the rich PL
spectra of TMD MLs by including indirect transitions associated
with electrons and holes in dissimilar valleys
\cite{Dery2015,Qiu2015,Echeverry2016,Selig2016,Selig2017} as
initially proposed by Dery and Song for combinations of electrons
in $\m{K}$ with empty VB states in $\m{K'}$ in tungsten-based MLs
\cite{Dery2015}. To this end we construct excitons by
forming an empty state in the upper valence sub-band at the
$\m{K}$ valley and the Coulomb-correlated electron at the $\m{K'}$
or, alternatively, at one of the $\m{Q}$-points. Note that
the hole state is formally associated with the time-reversal of
the unoccupied state in the valence band~\cite{Wang2017b}.
Neglecting the upper sub-band at the $\m{Q}$-points due to sizable
spin-orbit splittings of the order of $100$~meV
\cite{Kormanyos2015} and omitting electron-hole exchange for
simplicity (energy scale of a few meV), we obtain the exciton
spectrum shown schematically in Fig.~\ref{fig1}b. Two
zero-momentum configurations with both electron and hole at
$\m{K}$ correspond to the well studied spin-allowed and
spin-forbidden exciton ($\m{X}$ and $\m{D}$)
\cite{Zhang2017,Molas2017,Zhou2017,Wang2017a,Qiu2013,Echeverry2016}.

In addition to direct excitons, also excitons with finite center-of-mass momenta can be
constructed from electrons in valleys other than the
unoccupied state in $\m{K}$. They do not recombine
directly via photon emission but require the assistance of
acoustic or optical phonons.
We label these momentum-dark excitons with capital letters denoting
the electron valley with the subscript l (u) for
spin-like (spin-unlike) configurations of the electron and hole
spins (in electron spin notation). By neglecting
electron-hole exchange we obtain two pairs of degenerate states
with electrons and holes in $\m{K}$ ($\m{D}$ and $\m{K'_{l}}$ as
well as $\m{X}$ and $\m{K'_{u}}$), and degenerate spin-like and
spin-unlike $\m{Q}$-excitons with electrons in six inequivalent
$\m{Q}$-pockets. The energetic ordering in Fig.~\ref{fig1}b
corresponds to tungsten-based MLs. In the presence of time-reversal
symmetry, all states have their counterparts with the
unoccupied state at the $\m{K'}$-valley and reversed spin
orientation.

As the manifold of momentum-dark excitons, shown encaged in
Fig.~\ref{fig1}b, has no dipolar radiative pathways due to
momentum conservation constraints, the states do not appear
directly in PL or reflection spectroscopy. However, in analogy to indirect band-gap bulk
semiconductors such as silicon \cite{Chelikowsky1976} or hexagonal
boron nitride (hBN) \cite{Cassabois2016}, finite-momentum excitons
can decay radiatively via simultaneous emission of phonons. Such
decay channels, indicated schematically in Fig.~\ref{fig1}b and c
by colored arrows and enabled by acoustic and optical phonons as
well as higher order combinations of multi-phonon processes, will
give rise to phonon replicas of momentum-dark excitons in the PL
emission. Once the energy positions of all states are determined
from spectral decomposition, the splittings $\Delta_{\m{XD}}$ and
$\Delta_{\m{XQ}}$ are obtained as indicated in Fig.~\ref{fig1}c.


\begin{figure}[t]
\begin{center}
\includegraphics[scale=1.0]{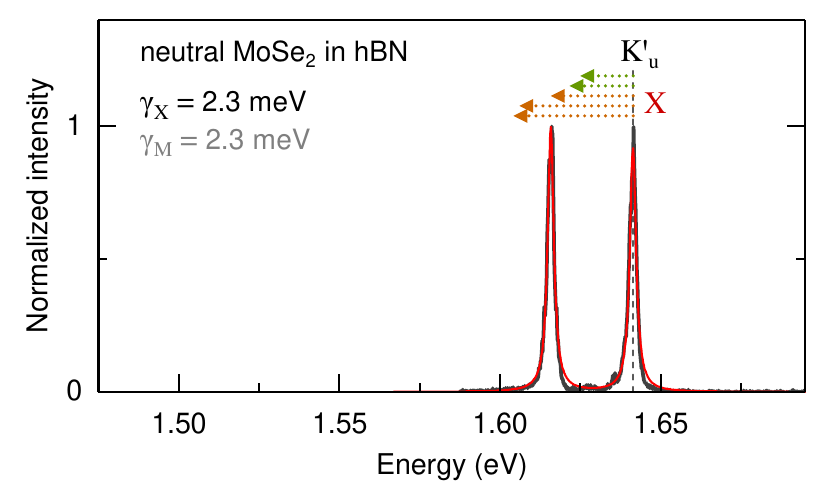}
\caption{ \textbf{Spectral decomposition of cryogenic
photoluminescence from monolayer MoSe$_\mathbf{2}$}. Basic
model fit (red solid line) with first-order phonon replicas of
momentum-dark $\m{K'_{u}}$ excitons resonant with the bright
exciton state $\m{X}$ in the absence of electron-hole exchange.
The best-fit energy position indicated by the dashed line
was obtained with $\gamma_{\m{X}}$ as fit parameter and
$\gamma_{\m{M}}$ set identical to $\gamma_{\m{X}}$. The green and
orange arrows indicate phonon sidebands of momentum-dark excitons
associated with acoustic and optical phonons with respective
energies taken from Ref.~\citenum{Jin2014}.
Free (fixed) fit parameters are given in the legends in black
(grey).}
\label{fig2}
\end{center}
\end{figure}
\textbf{Analysis on monolayer MoSe$_\mathbf{2}$ emission.} First,
we apply our analysis to ML MoSe$_2$ encapsulated in hBN with
active doping control. The cryogenic PL spectrum shown in
Fig.~\ref{fig2} features two bright PL peaks, commonly attributed
to the emission from neutral and charged excitons. In high
signal-to-noise differential reflectivity measurements in our
gated structure, however, no trion signature was detected in
addition to the solitary resonance of the neutral exciton (see
Supplementary Information) in contrast to doped samples
\cite{Chernikov2015}. We therefore argue that the intensive PL
peak $\sim 30$~meV below $\m{X}$ could also be interpreted as an
optical phonon sideband of the momentum-dark exciton state
$\m{K'_{u}}$ that we set resonant with the bright exciton by
neglecting electron-hole exchange. The respective acoustic
sidebands would then contribute weak yet finite PL in between the
two intensive peaks.

To obtain a model fit of the neutral ML MoSe$_2$ spectrum in the
framework of this analysis shown by the red solid line in
Fig.~\ref{fig2}, we modeled the ZPLs of resonant momentum-bright
and momentum-dark states $\m{X}$ and $\m{K'_{u}}$ by homogenously
broadened Lorentziants with the same full-width at half-maximum
linewidth $\gamma_\m{X}$. Moreover, we restricted the phonon
replicas of $\m{K'_{u}}$ to first-order processes. By taking the
corresponding phonon modes calculated in Ref.~\citenum{Jin2014}
(recapitulated in Table~S$1$ of the Supplementary
Information for convenience) with explicit phonon energies of
$16.6$, and $19.9$~meV for the TA and LA acoustic phonons, and
$35.5$, $37.4$, and $25.6$~meV for TO(E$'$), LO(E$'$) and
$\m{A}_1$ optical phonons available for the scattering of the
electron from the $\m{K'}$ into the $\m{K}$-valley, we allowed the
fitting procedure to determine the best-fit energy position
(indicated by the dashed line) and linewdith
$\gamma_\m{X}=2.3$~meV for the ZPL of $\m{X}$ and thus of
$\m{K'_{u}}$.

Remarkably, the correspondence between the spectrum and the model
fit in Fig.~\ref{fig2} was obtained with vanishing contributions
from TO and LO phonons, and thus the lower-energy peak can be
ascribed entirely to the $\m{A}_1$ optical sideband of $\m{K'}$.
Our analysis of ML MoSe$_2$ PL from a more disordered sample (see
Supplementary Information) indicates that both TO and LO phonons
as well as higher-order phonon processes can be activated in the
presence of disorder \cite{Cassabois2016a}. For phonon replicas to
be as intense in emission as the bright exciton emission in the PL
of neutral ML MoSe$_2$, long-lived population of dark states
without efficient decay channels must be present. Such population
can be provided by the reservoir of momentum-dark $\m{K}'$
excitons, or by momentum-dark $\m{Q}$ states if the value of
$28$~meV \cite{Jin2014} instead of the much higher prediction of
$137$~meV \cite{Kormanyos2015} is anticipated for the splitting
$\Delta_{\m{KQ}}$ in ML MoSe$_2$.

\begin{figure}[t]
\begin{center}
\includegraphics[scale=1.0]{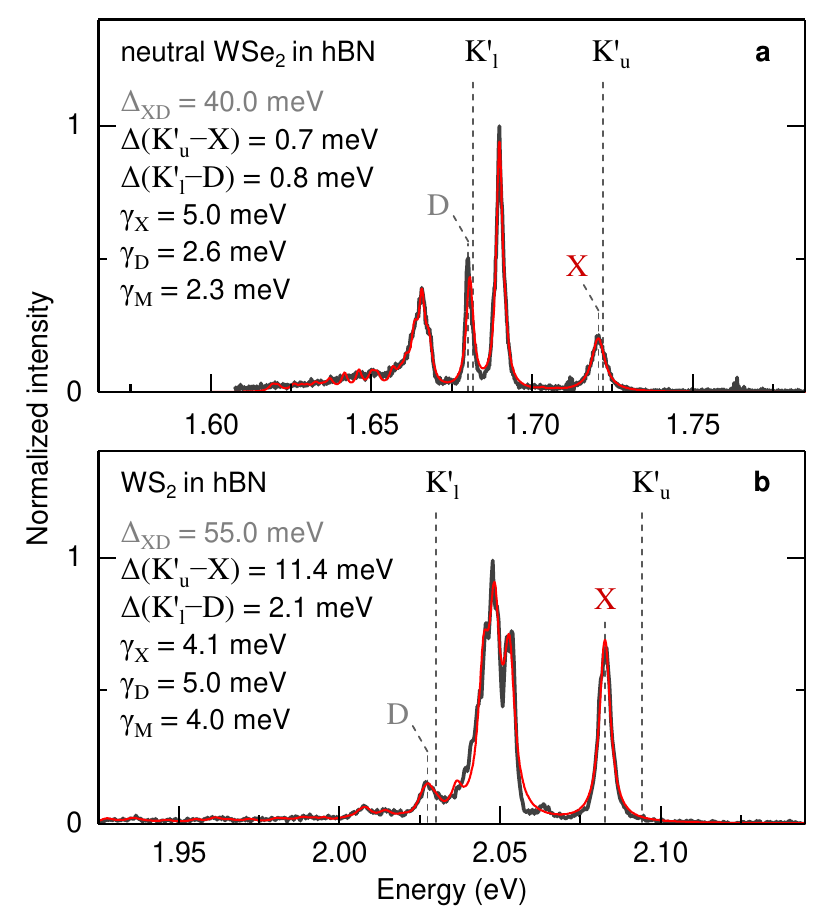}
\caption{\textbf{Spectral decomposition of cryogenic
photoluminescence from monolayer WSe$_\mathbf{2}$ and
WS$_\mathbf{2}$.} \textbf{a}, Refined model fit (red solid line)
to the spectrum of neutral monolayer WSe$_2$ including momentum-dark
state $\m{K'_{l}}$ resonant with $\m{D}$ at a fixed bright-dark
splitting $\Delta_{\m{XD}}=40$~meV and all other parameters
determined from the best fit. \textbf{b}, Same for monolayer
WS$_2$ without active charge control. Free (fixed) fit parameters
are given in the legends in black (grey).} \label{fig3}
\end{center}
\end{figure}

\textbf{Analysis of monolayer WSe$_\mathbf{2}$ emission.} The
analysis of the simple MoSe$_2$ emission has served as an
illustration of the possible involvement of phonon-assisted
recombination of momentum-dark excitons. In the next step we apply
our decomposition analysis to ML WSe$_2$ with a rich spectrum of
unidentified peaks \cite{Jones2013} as in Fig.~\ref{fig3}a
recorded on ML WSe$_2$ encapsulated in hBN and tuned to the point
of charge neutrality \cite{Courtade2017}. It features narrow
spectral lines characteristic of high-quality MLs with PL close to
the homogeneous limit \cite{Wang2017a,Ajayi2017,Cadiz2017} and we
assume a negligible contribution from trions, again based on the
absence of a trion resonance in high signal-to-noise reflectivity.
As discussed previously, the PL signatures of ML WSe$_2$ in
Fig.~\ref{fig3}a differ significantly from the PL of ML MoSe$_2$
in Fig.~\ref{fig2} because of the reversed ordering of
spin-polarized sub-bands in tungsten and molybdenum based
dichalcogenides. To model the PL spectrum of WSe$_2$, one has to
include the spin-forbidden exciton state $\m{D}$ red-shifted by
$40$~meV from the ZPL of the bright state $\m{X}$ in this specific
sample \cite{Courtade2017,Wang2017a}. In order to obtain the best
model fit shown as the red solid line in Fig.~\ref{fig3}a, we
allowed not only the phonon energies to vary around the values
given for ML $\m{WSe}_2$ in Ref.~\citenum{Jin2014} but also the
energies and linewidths of the Lorentzian ZPLs of $\m{D}$,
$\m{K'_{l}}$, $\m{K'_{u}}$, and $\m{X}$ states. Assuming similar
timescales for phonon-assisted decay and transform limited
broadening of momentum-dark states, a joint linewidth
$\gamma_{\m{M}}$ was used for both states.

The best-fit model spectrum of Fig.~\ref{fig3}a with up to
third-order processes was obtained with $\gamma_{\m{X}}=5.0$~meV
and comparable linewdiths of $\sim 2.5$~meV for both spin-forbidden and
momentum-dark states at the respective energy positions of the
ZPLs indicated by the dashed lines. The overall correspondence
between the measured spectrum and the model is again compelling.
It interprets the bright-most peak in between the bright and dark
exciton ZPLs as composed of optical phonon replicas of the
momentum-dark state $\m{K'_{u}}$, and the peak below $\m{D}$ as
acoustic sidebands of $\m{K'_{l}}$. We stress again that we do not
attribute here the peak 32~meV below the bright exciton to trion
emission for two reasons: First, the gated sample shows only one
solitary resonance of the neutral exciton without additional trion
features in differential reflectivity. Second, this sample tuned
into the \emph{n}-type regime exhibits a pronounced fine structure
splitting both in PL and reflectivity as a hallmark of negative
trions \cite{Courtade2017}. Without the ambiguity of unintentional
doping, the electron-hole exchange splittings
$\Delta(\m{K'_{u}}-\m{X})$ and $\Delta(\m{K'_{l}}-\m{D})$ are of
the order of a few meV.

\textbf{Analysis of monolayer WS$_\mathbf{2}$ emission.}
The analogous spectral decomposition was also carried out for ML WS$_2$
sandwiched in hBN without means of field-effect charge control.
The best fit to the PL spectrum of Fig.~\ref{fig3}b was obtained
according to the refined fitting procedure used for ML WSe$_2$ in
Fig.~\ref{fig3}a with a fixed bright-dark splitting of $55$~meV
derived from experiment \cite{Wang2017a} and similar values for
the linewdiths of momentum-bright and dark excitons in the range
of $4-5$~meV. It is worth pointing out the main similarities and
differences in the PL spectra for the two tungsten-based MLs. For
the WS$_2$ spectrum, only second-order processes were required
since the absolute energies are larger as compared to $\m{WSe}_2$
\cite{Jin2014}. Moreover, the phonon modes exhibit larger
splittings (see Supplementary Information, Table~S$1$). The LA-TA splitting at the $\m{K}$ point of
$\m{WS}_2$, for example, exceeds the value in $\m{WSe}_2$ by $\sim
4$~meV. More significantly, the optical phonon energies differ by
$\sim 15$~meV and up to $\sim 20$~meV at the $\Gamma$ and $\m{K}$
points, respectively.

Among the similar PL signatures is the weak peak below $\m{D}$ and
the intense peak between $\m{X}$ and $\m{D}$ with fine structure
due to the specific optical phonon spectrum of WS$_2$. Akin to
WSe$_2$, the former and the latter are assigned to acoustic and
optical replicas of momentum-dark states $\m{K'_{l}}$ and
$\m{K'_{u}}$, respectively. Surprisingly, best fit suggests an
exchange splitting of $\Delta(\m{K'_{u}}-\m{X})=11.4$~meV in
contrast to $2.7$~meV for WSe$_2$ in Fig.~\ref{fig3}a. The fit to
WS$_2$ PL requires a significant upshift of the state $\m{K'_{u}}$
in order to optimally accommodate the optical phonon sidebands
into the intense and complexly structured PL peak between $\m{X}$
and $\m{D}$. This could be an artefact of the non-quantified
contribution from trions in this sample, or indicate that the
set of involved momentum-dark excitons could be expanded by the
$\m{Q}$-exciton manifold as will be discussed in the next section.

\begin{figure}[t!]
\begin{center}
\includegraphics[scale=1.0]{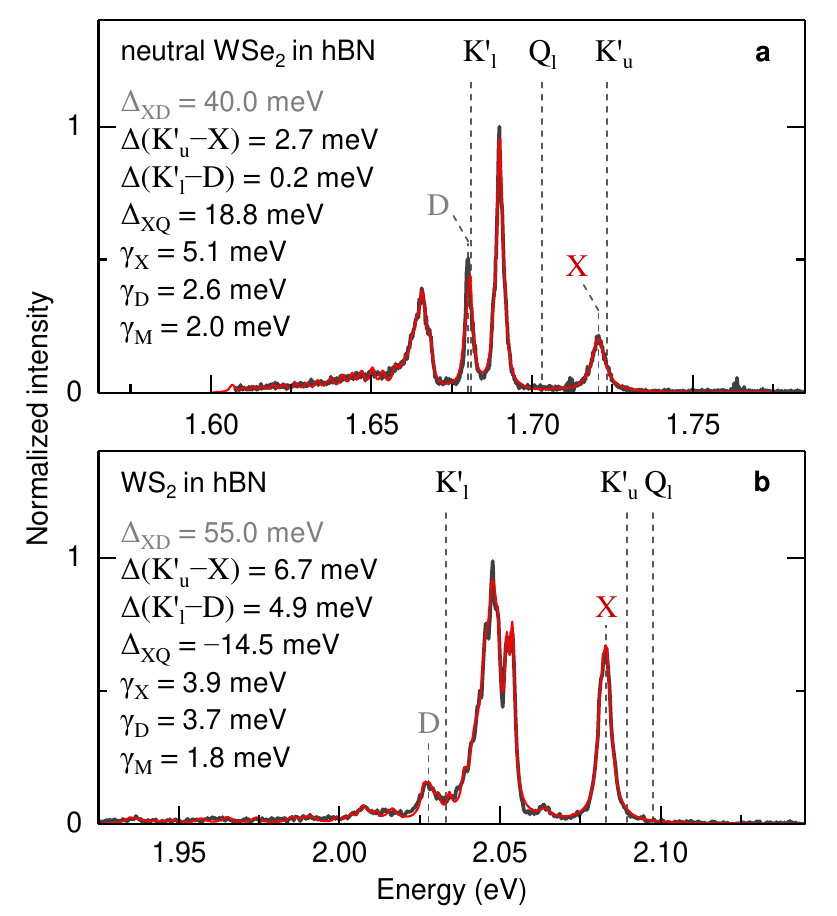}
\caption{ \textbf{Decomposition of monolayer WSe$_\mathbf{2}$ and
WS$_\mathbf{2}$ spectra including Q-momentum excitons.}
\textbf{a}, \textbf{b}, Same as Fig.~\ref{fig3}a, b but with
involvement of $\m{Q}$ exciton manifold placed in between the
states $\m{X}$ and $\m{D}$ in \textbf{a}, and above the state
$\m{X}$ in \textbf{b}. Free (fixed) fit parameters are given in
the legends in black (grey).} \label{fig4}
\end{center}
\end{figure}

\textbf{Role of $\m{\mathbf{Q}}$-momentum excitons in monolayer
WSe$_\mathbf{2}$ and WS$_\mathbf{2}$ emission.} For WSe$_2$ MLs
the $\m{Q}$-momentum excitons can play an important role, since
the $\m{Q}$-valley is in close proximity to the lowest CB minimum
at $\m{K}$ according to single-particle calculations
\cite{Jin2014,Kormanyos2015,Zhang2015a}. Excitonic corrections
have been predicted to reduce the energy level of $\m{Q}$-excitons
well below the energy of the lowest spin-forbidden state $\m{D}$ both
in WSe$_2$ and WS$_2$ MLs \cite{Selig2016,Selig2017,Malic2017}.
This, however, is in contradiction to the analysis developed so
far that explains the lowest-energy PL peak in terms of acoustic
replicas of the momentum-dark reservoir $\m{K'_{l}}$. Any deeper
momentum-dark state should exhibit large population with
pronounced PL phonon sidebands as in the case of bilayer WSe$_2$
with momentum-indirect band gap \cite{Lindlau2017}. The only two
remaining scenarios for the energy position of the $\m{Q}$-exciton
level is in between $\m{D}$ and $\m{X}$ or above $\m{X}$ (apart
from placing it in resonance with $\m{K'_{u}}$ or $\m{K'_{l}}$
with trivial implications).

The analysis of best fits shown in Fig.~\ref{fig4} suggests
that the first scenario is better suited to model the spectrum of
ML $\m{WSe}_2$. Before proceeding, we note that second- and
higher-order phonon-assisted processes were restricted to
combinations of multiple phonons with total phonon momentum of
$\m{Q}$ or $\m{K}$ depending on the respective initial valley of
the electron (see Supplementary Information for details).
For example, the scattering of the electron from the
$\m{Q}$-valley into the $\m{K}$-valley and subsequent emission of
an optical phonon would involve an LA or TA phonon at the
$\m{Q}$-point and a zero-momentum optical phonon at the
$\Gamma$-point of the first Brilloiun zone. With this approach to
the best-fit, the energy position of the $\m{Q_{l}}$ state in
Fig.~\ref{fig4}a is identified at $\Delta_{\m{XQ}}\simeq 19$~meV
below the bright exciton with marginal variations in other fit
parameters as compared to Fig.~\ref{fig3}a. The corresponding
energy level hierarchy would assign the bright-most PL peak now to
acoustic phonon replicas of the $\m{Q}$-exciton manifold with
contributions to the lowest-energy peak via optical sidebands.

In the case of ML WS$_2$ in Fig.~\ref{fig4}b, on the other
hand, the second scenario performed better. It adds an
explanation to the first weak PL peak below $\m{X}$ as an acoustic
sideband of $\m{Q_{l}}$ with its respective
optical sidebands merging into the most intense PL peak between
$\m{X}$ and $\m{D}$. Moreover, this configuration reduced the
conspicuously large exchange splitting between $\m{X}$ and
$\m{K'_{u}}$ found in the fit of
Fig.~\ref{fig3}b, and is at least qualitatively in line with
theoretical calculations that predict a small separation between
$\m{Q}$- and $\m{K}$-excitons in WS$_2$ rather than in WSe$_2$ MLs
\cite{Malic2017}.

Overall, within the suggested approach we find good
qualitative and satisfactory quantitative description of the
spectra. Its quantitative validity is limited by the assumption
of identical linewidths for all momentum-dark excitons which is
not necessarily the case since different phonon-assisted pathways
determine the effective lifetimes of momentum-dark excitons.
Moreover, as opposed to the inclusion of both in-plane and
out-of-plane optical phonon modes, we discarded the out-of-plane
acoustic phonon mode ZA. The experimental precision limited by the
spectral broadening even in best samples
\cite{Wang2017a,Cadiz2017,Ajayi2017} currently provides an upper
bound of a few meV on these effects.

Even with the current uncertainty in the values of exchange
interaction and the energetic splittings between the valleys, our
model highlights the importance of the role played by
momentum-dark excitons in elementary optical response of ML TMDs.
The conclusions are fully in line with the interpretation of
cryogenic spectra from bilayer WSe$_2$ \cite{Lindlau2017} and
MoSe$_2$-WSe$_2$ heterobilayers \cite{Forg2017}. Based on our
findings, further experimental work and more precise theoretical
calculations of the single-particle band structure and phonon
modes will finally consolidate a quantitative understanding of
excitons in TMD MLs. Placed into a broader perspective of
prevalent puzzles in TMD spectroscopy \cite{Koperski2017}, our
analysis provides sufficient guidelines for new interpretations.

{\bf Acknowledgments:} We thank G.~Cassabois for fruitful
discussions and P.~Altpeter and R.~Rath for assistance in the
clean-room. This work was funded by the Volkswagen Foundation, the
European Research Council (ERC) under the ERC grant agreement no.
$336749$, and the Deutsche Forschungsgemeinschaft (DFG) via the
Cluster of Excellence Nanosystems Initiative Munich (NIM). A.~H.
also acknowledges support from the Center for NanoScience (CeNS)
and LMUinnovativ. B.~U. thanks ERC grant agreement no. $306719$,
ITN Spin-NANO Marie Sklodowska-Curie grant agreement no. 676108,
C.~R. thanks ANR MoS2ValleyControl, Programme Investissements
d'Avenir ANR-$11$-IDEX-$0002$-$02$, reference ANR-$10$-
LABX-$0037$-NEXT for financial support. M.~M.~G. acknowledges
support from ILNACS CNRS-Ioffe and RFBR $17$-$52$-$16020$. X.~M.
also acknowledges the Institut Universitaire de France. K.~W. and
T.~T. acknowledge support from the Elemental Strategy Initiative
conducted by the MEXT, Japan and JSPS KAKENHI Grant Numbers
$\m{JP}26248061$, $\m{JP}15\m{K}21722$, and $\m{JP}25106006$.


\cleardoublepage

\setcounter{figure}{0} \setcounter{equation}{0} \setcounter{table}{0}
\renewcommand{\thetable}{S\arabic{table}}
\renewcommand{\figurename}{Figure~S}
\makeatletter
\def\fnum@figure{\figurename\thefigure}
\makeatother
\newcommand{\figref}[1]{Fig.~S\ref{#1}}

\section{Supplementary Information}

\subsection{{\bf Experimental setup}}

Cryogenic confocal PL spectroscopy studies were performed at
$3.1$~K in a closed-cycle cryostat (attocube systems,
attoDRY$1000$) or a helium dewar at $4.2$~K. The samples were
positioned with piezo-steppers and scanners (attocube systems,
ANP$101$ series and ANSxy$100$/lr) into the diffraction limited
spot of a low-temperature apochromatic objective with a numerical
aperture of $0.82$ (attocube systems, LT-APO/VISIR/$0.82$) and a
spot size of $0.6~\mu$m. A He-Ne laser at $632$~nm, a continuous
wave green laser at $532$~nm or a supercontinuum laser (NKT
Photonics, SuperK EXW-$12$) operated at $532$~nm with a spectral
width of $6$~nm were used to excite the PL. A monochromator (PI,
Acton SP-$2558$) equipped with a nitrogen-cooled silicon CCD (PI,
Spec-$10$:$100$BR/LN) was used to detect PL with a spectral
resolution of $0.4$~meV.

\subsection{{\bf Gated samples}}

We have fabricated van der Waals heterostructures by mechanical
exfoliation from commercially available bulk crystals and very
high quality hBN crystals~\cite{Taniguchi:2007a}. A first layer of
hBN was mechanically exfoliated and transferred onto a SiO$_2$ (90~nm)/Si substrate using PDMS stamping \cite{Gomez:2014a}. The
deposition of the subsequent ML and the second hBN capping
layer was obtained by repeating this procedure to complete the
full stack. We also transferred a thin graphite flake between the
top surface of the ML and a Au pre-patterned electrode.
Carrier concentration was varied by applying a bias between this
electrode and the \emph{p}-doped Si substrate (back gate).

\subsection{{\bf Photoluminescence and differential reflectivity}}

\begin{figure}[h!]
\begin{center}
\includegraphics[scale=1.0]{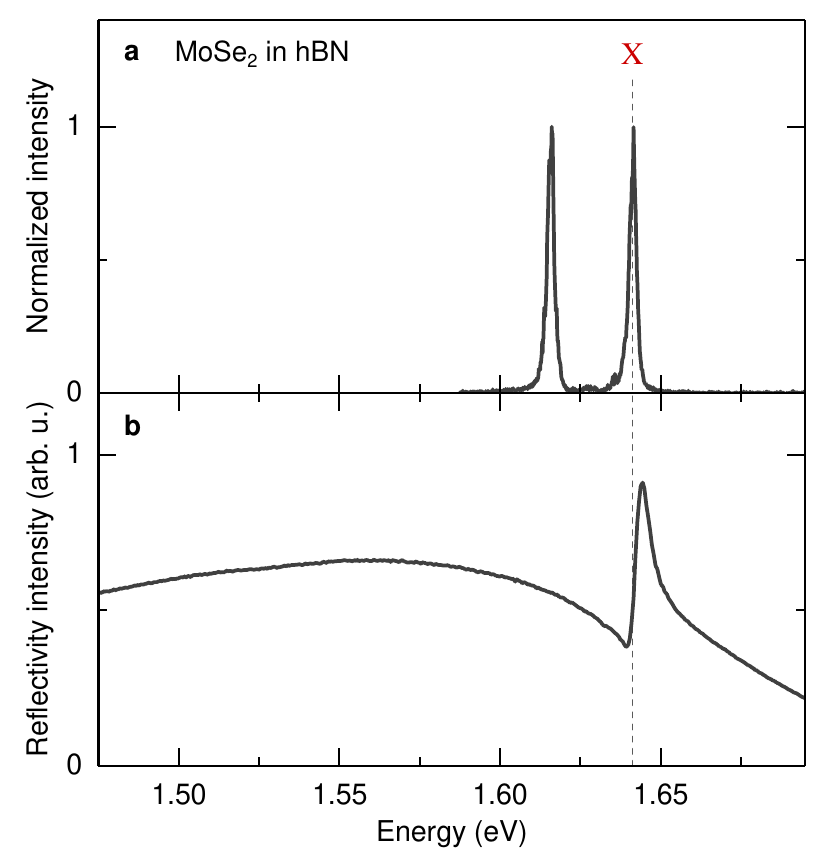}
\caption{\textbf{a}, Photoluminescence spectrum of monolayer
MoSe$_2$ adopted from Fig.~\ref{fig2} of the main text.
\textbf{b}, Corresponding reflectivity at $V_g=+10$~V.
Note the absence of trion-related features.} \label{S2_MoSe2-DR}
\end{center}
\end{figure}

\begin{figure}[h!]
\begin{center}
\includegraphics[scale=1.0]{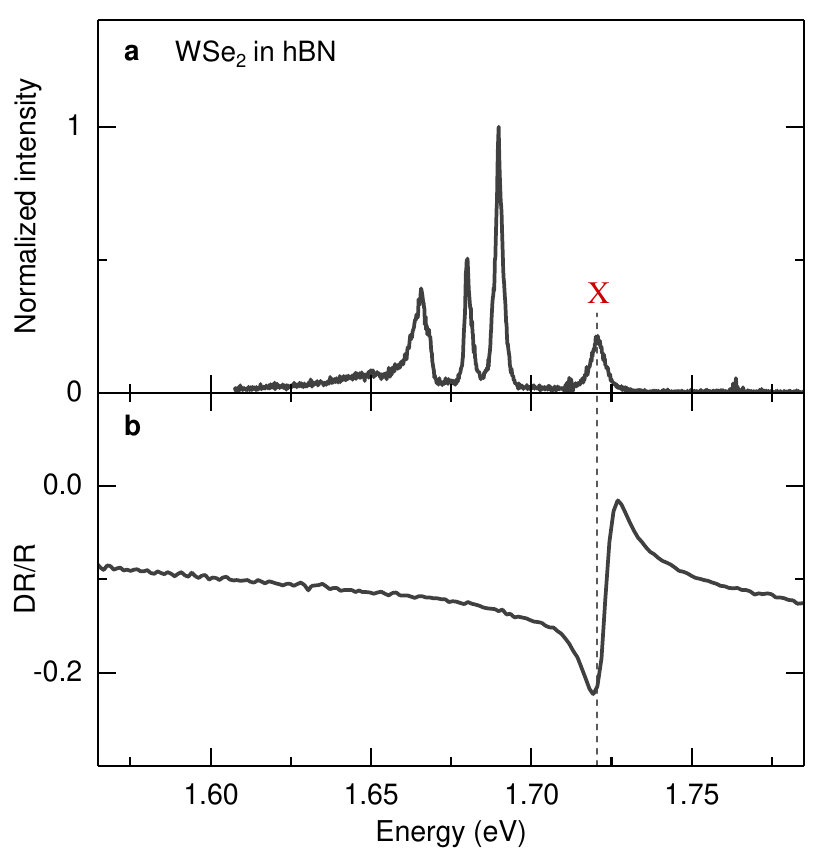}
\caption{\textbf{a}, Photoluminescence spectrum of monolayer
WSe$_2$ adopted from Fig.~\ref{fig3}a of the main text.
\textbf{b}, Corresponding differential reflectivity. Note the absence of trion-related features.}
\label{S3_WSe2-DR}
\end{center}
\end{figure}

The charge doping level of ML MoSe$_2$ and WSe$_2$ was controlled
with field-effect devices described above and monitored with
differential reflectivity (DR). The PL and reflectivity spectra of ML
MoSe$_2$ at the charge neutrality point are shown in
Fig.~S\ref{S2_MoSe2-DR}a and b, respectively as in \cite{Courtade2017}. The reflectivity spectrum shows
no signature of trion absorption. The respective PL and DR spectra
of ML WSe$_2$ at the charge neutrality point are shown in
Fig.~S\ref{S3_WSe2-DR}a and b, respectively. As for ML MoSe$_2$,
the DR spectrum shows no signature of trion absorption.

\begin{figure}[t!]
\begin{center}
\includegraphics[scale=1.0]{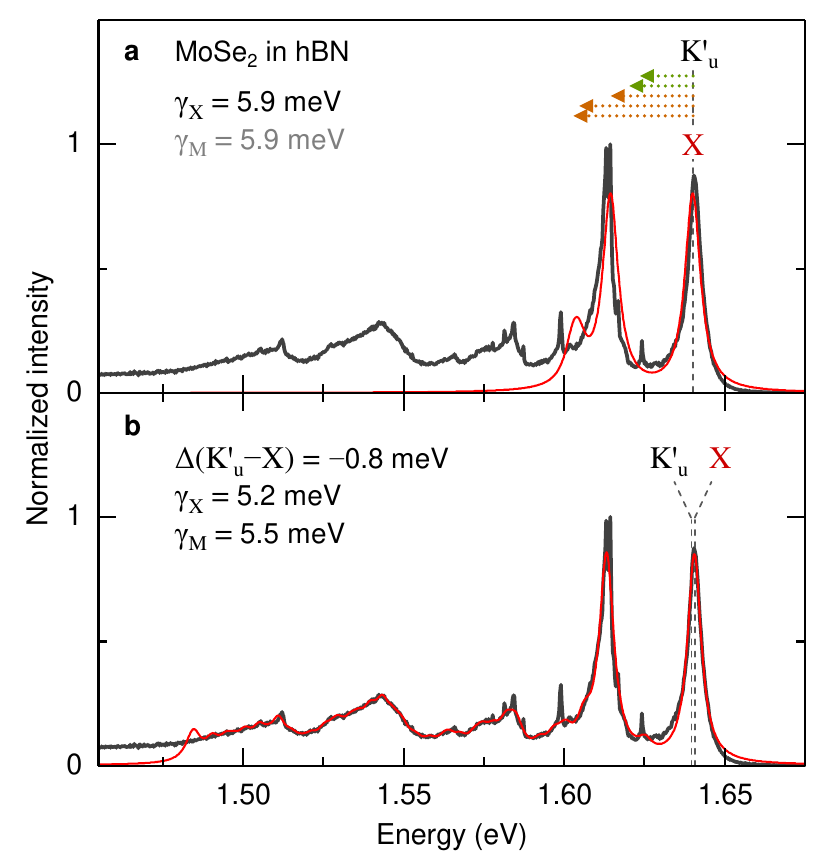}
\caption{Spectral decomposition of cryogenic photoluminescence
from monolayer MoSe$_\mathbf{2}$ without active control of charge
doping. \textbf{a}, Basic model fit (red solid line) with
first-order phonon replicas of momentum-dark $\m{K'_{u}}$ excitons
resonant with the bright exciton state $\m{X}$ in the absence of
electron-hole exchange. The best-fit energy position indicated by
the dashed line was obtained with $\gamma_{\m{X}}$ as fit
parameter and $\gamma_{\m{M}}$ set identical to $\gamma_{\m{X}}$.
The green and orange arrows indicate phonon sidebands of
momentum-dark excitons associated with acoustic and optical
phonons with respective energies taken from
Ref.~\cite{Jin2014}. \textbf{b}, Refined model fit (red solid
line) with variable energy positions and linewdiths of $\m{X}$ and
$\m{K'_{u}}$ states and up to fourth order phonon replicas with
variable phonon energies bound by $\pm 2$~meV around the values of
Ref.~\cite{Jin2014}. Free (fixed) fit parameters are given in
the legends in black (grey).} \label{S1_MoSe2-PL}
\end{center}
\end{figure}

We also applied our analysis to ML MoSe$_2$ encapsulated in hBN
without active doping control. As in Fig.~\ref{fig2} of the main
text, the PL spectrum in Fig.~S\ref{S1_MoSe2-PL} features two
bright PL peaks. In addition, the PL exhibits an extended red wing
with some structure commonly ascribed to localized excitons in
potentials of unintentional disorder. Here, we assume that the
intensive PL peak $\sim 30$~meV below $\m{X}$ is not related to
trions but is instead composed of optical phonon sidebands of the
momentum-dark exciton state $\m{K'_{u}}$ resonant with the bright
exciton in the absence of electron-hole exchange.

The model fit to the ML MoSe$_2$ spectrum of
Fig.~S\ref{S1_MoSe2-PL}a was obtained with ZPLs of momentum-bright
and momentum-dark excitons modeled by homogenously broadened
Lorentziants at the same energy and with the same full-width at
half-maximum linewidth $\gamma_\m{X}$. Analogous to Fig.~\ref{fig2}
of the main text, first-order scattering processes by acoustic and
optical phonons with energies from Ref.~\citenum{Jin2014} yield
the two peaks as the main PL features with $\gamma_\m{X}=5.9$~meV
and best-fit energy positions indicated by the dashed lines.

To improve the fit up to the striking correspondence with the
spectrum in Fig.~S\ref{S1_MoSe2-PL}b, we allowed the phonon
energies to vary by $\pm 2$~meV around their theoretical values.
Such small variation of phonon energies account for
sample-to-sample variations in the dielectric environment or
strain and are well within the range of quantitative observations
with Raman spectroscopy \cite{Zhang2015Raman}. Moreover, we
included phonon processes of up to fourth order (the cut-off to the
model spectrum around $1.48$~eV is because processes beyond fourth
order were truncated), and allowed the energy positions and the
linewidths to vary for both $\m{X}$ and $\m{K'_{u}}$ states.
Remarkably, all intricate features of the PL spectrum are well
reproduced by the model fit without significant changes to the ZPL
energies and linewidths, and with higher-order phonon processes
improving the correspondence between the fit and the intricate
spectral details of the extended red tail of the PL spectrum. The
bright peak below $\m{X}$ is interpreted as composed of optical
phonon sidebands of the momentum-dark state $\m{K'_{u}}$ that also
gives rise to broad lower-energy PL peaks via its higher-order
phonon replicas. In contrast, the emission from disorder-localized
excitons \cite{Branny2016}, characterized by narrow spectral features in
Fig.~S\ref{S1_MoSe2-PL}b, is not captured by the present model.

\begin{table}[h!]
\centering
        \small{
        \begin{tabular}{c|ccc|ccc|ccc}
        \toprule
                & \multicolumn{3}{|c|}{MoSe$_2$} &  \multicolumn{3}{|c|}{WS$_2$} & \multicolumn{3}{|c}{WSe$_2$} \\
        \hline
                Mode & $\Gamma$ & K & Q & $\Gamma$ & K & Q & $\Gamma$ & K & Q \\
                \hline
        TA    &$0$&$16.6$&$13.3$  & $0$&$17.4$&$15.9$ & $0$&$15.6$& $11.6$ \\
        LA    &$0$&$19.9$&$16.9$  & $0$&$23.6$&$19.5$ & $0$&$18.0$&$14.3$ \\
        TO(E$'$)  &\textcolor[rgb]{0.6,0.6,0.6}{$36.1$}&$35.5$&$36.4$ & \textcolor[rgb]{0.6,0.6,0.6}{$44.4$}&$43.8$&$45.3$ & \textcolor[rgb]{0.6,0.6,0.6}{$30.5$}&$26.7$&$27.3$ \\
                LO(E$'$)  &$36.6$&$37.4$&$37.5$ & $44.2$&$43.2$&$42.3$ & $30.8$&$31.5$&$32.5$ \\
                A$_1$     &$30.3$&$25.6$&$27.1$ & $51.8$&$48.0$&$50.0$ & \textcolor[rgb]{0.6,0.6,0.6}{$30.8$}&$31.0$&$30.4$ \\
       \hline\hline
        \end{tabular}
        }
\caption{Phonon mode energies at the high-symmetry points of
the first Brillouin zone for monolayer MoSe$_2$, WS$_2$, and
WSe$_2$ used in the model fits. Higher order scattering processes
with phonon energies equal to the energy of LO(E$'$) within
$1$~meV (listed in the table in grey) were discarded from our
analysis for simplicity. All energies are given in meV and
reproduced from Ref.~\citenum{Jin2014}.} \label{E_phonons}
\end{table}

\subsection{{\bf Group theory analysis}}

The symmetry of the Q-point is $C_{s}$ with only two symmetry
operations: identity and horizontal plane reflection. There are
two irreducible (vector) representations of this group,
namely A$'$ (invariant) and A$''$ (z-coordinate, i.e. normal to
the reflection plane). The intersection of the two symmetry groups
$C_{s}$ (Q-point) and $C_{3h}$ (K-point) is $C_{s}$. The
conduction band both at the Q- and K-points corresponds to the
A$'$ representation and, hence, transitions are possible via
phonons with the same symmetry A$'$ (these modes are symmetric
under $z\rightarrow-z$). All phonon modes in question indeed
correspond to this representation: acoustic E$'$ at the
$\Gamma$-point corresponds to A$'$ at the Q-point, optical E$'$ at
the $\Gamma$-point corresponds to A$'$ at the Q-point, and optical
A$_1'$ at the $\Gamma$-point corresponds to A$'$ at the Q-point.
Combinations of these phonons are also allowed (provided that
momentum conservation is fulfilled). With account for spin-orbit
interaction all other phonons (asymmetric for $z\rightarrow-z$)
are also active in $\m{K}$ to $\m{Q}$ transitions. In order
to account for spin-orbit effects we need to also consider the
spinor representations $\Gamma_3$ (spin-up, $\uparrow$) and
$\Gamma_4$ (spin-down, $\downarrow$):
$\Gamma_3=\m{A}'\times\uparrow = \m{A}''\times\downarrow$,
$\Gamma_4 = \m{A}''\times \uparrow = \m{A}'\times \downarrow$.
Thus, spin-up and spin-down states can be mixed with Bloch
functions of different orbital symmetry and odd in $z \to
-z$ phonon modes can enable spin-flip transfer between K and Q
valleys.

Transitions between the K- and K$'$-points can be considered
analogously. To that end, we note that both elements of the
wavevector group $C_{3h}$ leave the K and K$'$ valleys intact. The
orbital Bloch functions of the conduction bands belong to the
E$'(1)$ ($\sim x+\mathrm i y$) and E$'(2)$ ($\sim x-\mathrm i y$)
irreducible representations of the $C_{3h}$ point group. The
transitions between the conduction band states are enabled by
phonon modes with E$'$ symmetry. In contrast, the orbital Bloch
functions of the valence band transform according to the A$'$
(invariant) irreducible representation. Thus, hole intervalley
scattering is provided by the fully symmetric A$'$ modes.

In the following we briefly address the activation of the
spin-unlike $\m{K'_{u}}$ excitons. In accordance with the symmetry
analysis these states form basis irreducible representations
E$''(1)$ and E$''(2)$ of the $C_{3h}$ point symmetry group
transforming as $(x\pm i y) z$. Formally, these states can be
transferred to the optically active ones by phonons with A$''$
symmetry (transforming like $z$) or by E$''$ phonons. Moreover,
the interaction with E$'$ phonons converts these excitons into
$z$-polarized states. If the mirror symmetry is distorted, the
$z\to-z$ operation (together with $S_3$ mirror rotation) should be
excluded from the $C_{3h}$ point symmetry group and the
representations E$''$ (odd in $z\to -z$) and E$'$ (even at $z\to
-z$) cannot be formally distinguished. Hence, in experiments
involving different dielectric environments at the top and the
bottom of TMD MLs, A$'$ and E$''$ phonons may enable activation of
spin-unlike $\m{K'_u}$ excitons.

\end{document}